\documentclass[a4paper,twoside,twocolumn,english,prb,floatfix,showpacs,preprintnumbers,superscriptaddress,amsmath,amssymb]{revtex4}
\usepackage[T1]{fontenc}
\usepackage[latin1]{inputenc}
\usepackage{amsmath}
\usepackage{graphicx}
\usepackage{amssymb}

\makeatletter

\newcommand{\noun}[1]{\textsc{#1}}

\usepackage{babel}
\makeatother
\begin{document}

\title{Tilt-angle landscapes and temperature dependence of the conductance
in \\
biphenyl-dithiol single-molecule junctions}

\author{F.~Pauly}

\email{Fabian.Pauly@kit.edu}

\affiliation{Institut f\"ur Theoretische Festk\"orperphysik and DFG-Center for
Functional Nanostructures, Universit\"at Karlsruhe, D-76128 Karlsruhe,
Germany}

\author{J.~K.~Viljas}

\affiliation{Institut f\"ur Theoretische Festk\"orperphysik and DFG-Center for
Functional Nanostructures, Universit\"at Karlsruhe, D-76128 Karlsruhe,
Germany}

\affiliation{Forschungszentrum Karlsruhe, Institut f\"ur Nanotechnologie, D-76021
Karlsruhe, Germany}

\author{J.~C.~Cuevas}

\affiliation{Departamento de F\'{\i}sica Te\'orica de la Materia Condensada,
Universidad Aut\'onoma de Madrid, E-28049 Madrid, Spain}

\affiliation{Institut f\"ur Theoretische Festk\"orperphysik and DFG-Center for
Functional Nanostructures, Universit\"at Karlsruhe, D-76128 Karlsruhe,
Germany}

\affiliation{Forschungszentrum Karlsruhe, Institut f\"ur Nanotechnologie, D-76021
Karlsruhe, Germany}

\author{Gerd Sch\"on}

\affiliation{Institut f\"ur Theoretische Festk\"orperphysik and DFG-Center for
Functional Nanostructures, Universit\"at Karlsruhe, D-76128 Karlsruhe,
Germany}

\affiliation{Forschungszentrum Karlsruhe, Institut f\"ur Nanotechnologie, D-76021
Karlsruhe, Germany}

\date{\today}


\begin{abstract}
Using a density-functional-based transport method we study the conduction
properties of several biphenyl-derived dithiol (BPDDT) molecules wired
to gold electrodes. The BPDDT molecules differ in their side groups,
which control the degree of conjugation of the $\pi$-electron system.
We have analyzed the dependence of the low-bias zero-temperature conductance
on the tilt angle $\varphi$ between the two phenyl ring units, and
find that it follows closely a $\cos^{2}\varphi$ law, as expected
from an effective $\pi$-orbital coupling model. We show that the
tilting of the phenyl rings results in a decrease of the zero-temperature
conductance by roughly two orders of magnitude, when going from a
planar conformation to a configuration in which the rings are perpendicular.
In addition we demonstrate that the side groups, apart from determining
$\varphi$, have no influence on the conductance. All this is in agreement
with the recent experiment by Venkataraman \emph{et al.}~{[}Nature
\textbf{442}, 904 (2006){]}. Finally, we study the temperature dependence
of both the conductance and its fluctuations and find qualitative
differences between the examined molecules. In this analysis we consider
two contributions to the temperature behavior, one coming from the
Fermi functions and the other one from a thermal average over different
contact configurations. We illustrate that the fluctuations of the
conductance due to temperature-induced changes in the geometric structure
of the molecule can be reduced by an appropriate design.
\end{abstract}
\maketitle

\section{Introduction}

In atomic-scale conductors the precise positions of the atoms have
a decisive influence on the electronic transport properties.\cite{Agrait:RhysRep2003,Hu:PRL2005,Pauly:PRB2006}
In the case of metal-molecule-metal contacts the importance of such
details often complicates the reproducibility of the experimental
results.\cite{Reichert:PRL2002,Venkataraman:NanoLett2006,Ulrich:JPhysChemB2006}
For this reason statistical analyses of the experimental data, such
as conductance histograms, have become indispensable for exploring
the charge-transport characteristics of single-molecule contacts.\cite{Cui:Science2001,Smit:Nature2002,Xu:Science2003,Venkataraman:NanoLett2006,Ulrich:JPhysChemB2006,Li:Nanotechnology2007}

Recently, making use of conductance histograms, Venkataraman \emph{et
al.~}were able to reveal the influence of molecular conformation
on the conductance of single-molecule contacts.\cite{Venkataraman:Nature2006}
In their experiments, these authors investigated biphenyl-derived
molecules, where different side groups were used to control the tilt
angle $\varphi$ between two phenyl rings. Thereby, the extent of
the delocalized $\pi$-electron system of the molecules could be varied.
They found that the conductance exhibited a characteristic $\cos^{2}\varphi$
behavior, as expected from a simple effective $\pi$-orbital coupling
model.\cite{Woitellier:ChemPhys1989,Samanta:PRB1996}

\begin{figure}
\begin{center}\includegraphics[%
  width=0.7\columnwidth,
  keepaspectratio]{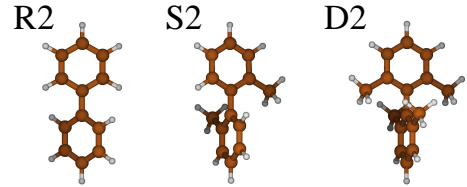}\end{center}

\caption{\label{cap:R2S2D2}(Color online) Biphenyl molecules R2, S2, and
D2. For S2 one hydrogen atom in the ortho position with respect to
the ring-connecting carbons in each phenyl ring of R2 is replaced
by a methyl group. For D2 also the second ortho-positioned hydrogen
is substituted by a methyl group.}
\end{figure}

Motivated by the experiment of Ref.~\onlinecite{Venkataraman:Nature2006},
we analyze theoretically the charge-transport properties of three
different biphenyl-derived dithiol (BPDDT) molecules connected to
gold electrodes. For simplicity, we refer to these molecules as R2,
S2, and D2 (Fig.~\ref{cap:R2S2D2}).%
 While R2 is the conventional biphenyl, the other two molecules, 2,2'-dimethyl-biphenyl
(S2) and 2,6,2',6'-tetramethyl-biphenyl (D2), differ from R2 by the
incorporation of one or two methyl groups in the ortho position with
respect to the ring-connecting carbon atoms. 

As an important difference to Ref.~\onlinecite{Venkataraman:Nature2006},
we bind the molecules R2, S2, and D2 to gold using thiol end groups
instead of amino groups. It has recently been shown that amino groups
are better suited to establish a reproducible contact of a molecule
to a gold electrode.\cite{Venkataraman:NanoLett2006,Ulrich:JPhysChemB2006}
These findings have been explained by the less directional character
of the amine-gold bond as compared to the thiol-gold linkage. Nevertheless,
thiol groups remain a frequent choice to establish the electrode-molecule
contact,\cite{Cui:Science2001,Reichert:PRL2002,Loertscher:Small2006,Li:Nanotechnology2007}
and the molecules R2, S2, and D2 with acetyl-protected sulfur end
groups have recently been synthesized.\cite{Shaporenko:JChemPhys2006}
Besides, it is the internal structure of the molecules that is most
important for the charge-transport characteristics discussed below.

In this work we investigate the effect of the degree of $\pi$ conjugation
on the conductance of BPDDT molecules connected to gold electrodes.
For this purpose, we describe the electronic structure of the single-molecule
contacts at the level of density functional theory (DFT). We demonstrate
that, in agreement with the experiments of Ref.~\onlinecite{Venkataraman:Nature2006},
a $\cos^{2}\varphi$ behavior of the low-bias zero-temperature conductance
is obtained. This behavior is by and large independent of the methyl
side groups introduced. We find that the breaking of the conjugation
reduces the zero-temperature conductance by roughly two orders of
magnitude. In addition, we study the temperature dependence of the
conductance for all our molecules. For this we take two contributions
into account. The first one comes from the broadening of the Fermi
functions of the leads, the other one from a thermal average over
different geometric configurations. In our analysis we observe that
S2 and D2 exhibit a monotonously increasing conductance as a function
of temperature, while for R2 the temperature dependence is non-monotonous.
Finally, we demonstrate that the temperature-fluctuations of the conductance
of single-molecule contacts can be reduced by an appropriate design
of the geometric structure of a molecule. This design should aim at
stabilizing the molecule with respect to the internal degrees of freedom
that are most relevant for its conduction properties. Due to the elimination
of uncertainties about the molecule's internal structure in a contact,
a more reliable comparison of experimental and theoretical results
on the charge-transport characteristics of single-molecule junctions
can be expected. 

The rest of the paper is organized as follows. In Sec.~\ref{sec:Theoretical-Model}
we outline the methods used to compute the electronic structure, geometry,
and conductance of the molecular contacts discussed below. Sec.~\ref{sec:Results-and-Discussion}
is devoted to the discussion of the results for the conductance of
the three BPDDT molecules, in particular their tilt-angle dependence
and temperature behavior. Technical details on these issues are deferred
to Apps.~\ref{sec:App-GT} and \ref{sec:App-eff-pi-coupl-model}.
Finally, we summarize our results in the conclusions, Sec.~\ref{sec:Conclusions}.

\section{Theoretical Model\label{sec:Theoretical-Model}}

In this section we present the methods applied in our work. These
include the procedures for computing the electronic structure, the
contact geometries, and the conduction properties of the molecular
contacts. For further details on our method we refer the reader to
Refs.~\onlinecite{Pauly:PhD2007,Wohlthat:arXiv2007,Viljas:arXiv2007}.

\subsection{Electronic structure and contact geometries}

For the determination of the electronic structure we employ DFT as
implemented in the RI-DFT module of the quantum chemistry package
\noun{Turbomole} \noun{v5.7} (Refs.~\onlinecite{Ahlrichs:ChemPhysLett1989,Eichkorn:ChemPhysLett1995}).
All our calculations, including the electrode description, are done
within \noun{Turbomole}'s standard Gaussian basis set, which is
of split-valence quality with polarization functions on all non-hydrogen
atoms.\cite{Schaefer:JChemPhys1992,Eichkorn:ChemPhysLett1995,Eichkorn:TheorChemAcc1992}
As the exchange-correlation functional we use BP86 (Refs.~\onlinecite{Becke:PhysRevA1988,Perdew:PhysRevB1986,Vosko:CanJPhys1980}).
All calculations were performed in a closed-shell formalism, and total
energies were converged to a precision of better than $10^{-6}$ atomic
units.

\begin{figure}
\begin{center}\includegraphics[%
  width=0.88\columnwidth,
  keepaspectratio]{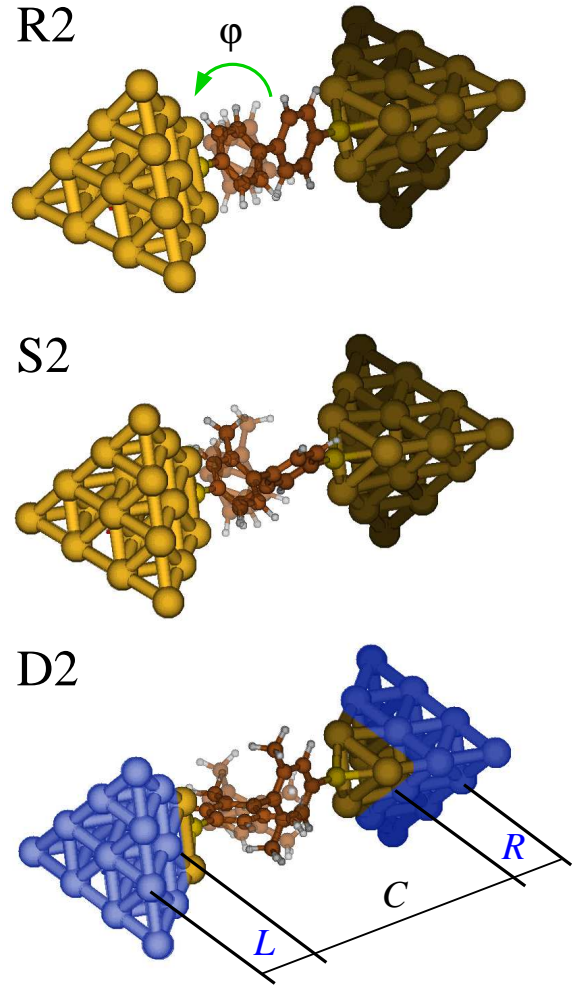}\end{center}

\caption{\label{cap:Au-R2S2D2-Au}(Color online) Molecules R2, S2, and D2
contacted at both ends to Au {[}111{]} electrodes via a sulfur atom
in a three-fold binding position. Faintly overlayed on the ground-state
structure of the single-molecule contacts are geometries, where the
left ring is rotated to $\varphi=\varphi_{0}\pm30°$. Here $\varphi$,
as indicated for R2, is the tilt angle between the planes of the two
phenyl rings, and $\varphi_{0}$ is its ground-state value. The division
of the junctions into the left ($L$), central ($C$), and right ($R$)
regions, of relevance in the conductance calculations, is also shown. }
\end{figure}

Our contact geometries are displayed in Fig.~\ref{cap:Au-R2S2D2-Au}.%
 For their determination we calculate at first the geo\-me\-tric
structure of a gold (Au) {[}111{]} pyramid with a thiolated benzene
molecule on top. This pyramid consists of three layers of Au with
3, 6, and 10 atoms. The tip atom of the pyramid is missing so that
the sulfur (S) atom of the benzene binds to a threefold hollow site
on top of the Au structure. We relax all atomic positions except for
the layers containing 6 and 10 atoms, which are kept fixed. In particular,
their lattice constant is set to $4.08$~$\textrm{Å}$, the experimental
bulk atomic distance of gold. Next, we compute the geometry of the
biphenyl molecules R2, S2, and D2 of Fig.~\ref{cap:R2S2D2} with
the hydrogen atoms in the position 4 and 4' substituted by sulfur
atoms, which are bonded to a single gold atom, respectively. For each
of these molecules we replace the single Au atoms on each side with
the previously mentioned Au {[}111{]} pyramids, where the thiolated
benzene molecule has been removed. In this process we take care that
the sulfur atoms of the biphenyl molecules are in the old positions
of the sulfur atoms of the thiolated benzene on top of the Au pyramids,
and that the $\mathrm{S-S}$ molecular axis and the {[}111{]} direction
are aligned. The ground-state contact geometry is subsequently obtained
by relaxing the complete structure, where we keep only the terminal
two gold layers on each side fixed (atoms shown in blue in Fig.~\ref{cap:Au-R2S2D2-Au}).
As above, the lattice constant in these layers is $4.08$~$\textrm{Å}$.
The relaxations thus include all atoms of the molecule, two sulfur
atoms, and six Au atoms, three on each side. In the determination
of the contact geometries we let the maximum norm of the Cartesian
gradient decay to values below $10^{-4}$ atomic units.

\subsection{Transmission function}

To compute the charge transport we apply a method based on standard
Green's function techniques and the Landauer formula expressed in
a local nonorthogonal basis.\cite{Xue:ChemPhys2002,Viljas:PRB2005,Wohlthat:arXiv2007}
The local basis allows us to partition the basis states into left
($L$), central ($C$), and right ($R$) parts, according to a division
of the contact geometry. Thus the Hamiltonian (or single-particle
Fock) matrix $\boldsymbol{H}$, and analogously the overlap matrix
$\boldsymbol{S}$, can be written in the block form\[
\boldsymbol{H}=\left(\begin{array}{ccc}
\boldsymbol{H}_{LL} & \boldsymbol{H}_{LC} & \boldsymbol{0}\\
\boldsymbol{H}_{CL} & \boldsymbol{H}_{CC} & \boldsymbol{H}_{CR}\\
\boldsymbol{0} & \boldsymbol{H}_{RC} & \boldsymbol{H}_{RR}\end{array}\right).\]
Within the Green's function method the energy-dependent transmission
$\tau(E)$ is expressed as\cite{Datta:Cambridge1995}\begin{equation}
\tau(E)=\mathrm{Tr}\left[\boldsymbol{\Gamma}_{L}\boldsymbol{G}_{CC}^{r}\boldsymbol{\Gamma}_{R}\boldsymbol{G}_{CC}^{a}\right],\label{eq:TE}\end{equation}
with the Green's functions\[
\boldsymbol{G}_{CC}^{r}(E)=\left[E\boldsymbol{S}_{CC}-\boldsymbol{H}_{CC}-\boldsymbol{\Sigma}_{L}^{r}(E)-\boldsymbol{\Sigma}_{R}^{r}(E)\right]^{-1}\]
and $\boldsymbol{G}_{CC}^{a}=\left[\boldsymbol{G}_{CC}^{r}\right]^{\dagger}$,
the self energies\begin{equation}
\boldsymbol{\Sigma}_{X}^{r}(E)=\left(\boldsymbol{H}_{CX}-E\boldsymbol{S}_{CX}\right)\boldsymbol{g}_{XX}^{r}(E)\left(\boldsymbol{H}_{XC}-E\boldsymbol{S}_{XC}\right),\label{eq:Simga_X}\end{equation}
the scattering rate matrices $\boldsymbol{\Gamma}_{X}(E)=-2\mathrm{Im}\left[\boldsymbol{\Sigma}_{X}^{r}(E)\right]$,
and the electrode Green's function $\boldsymbol{g}_{XX}^{r}$, where
$X=L,\, R$.

In Fig.~\ref{cap:Au-R2S2D2-Au} we show how we divide our contacts
into the $L$, $C$, and $R$ regions. In this way we obtain $\boldsymbol{H}_{CC}$
and $\boldsymbol{S}_{CC}$ for the $C$ region, which consists of
the BPDDT molecule and three gold atoms on each side of the junction.
The $L$ and $R$ regions are made up of the two terminal gold layers
on each side of the gold pyramids (blue shaded atoms of Fig.~\ref{cap:Au-R2S2D2-Au}),
which have been kept fixed to bulk atomic distances in the geometry
relaxations. The matrices $\boldsymbol{H}_{CX}$ and $\boldsymbol{S}_{CX}$,
extracted from these finite contact geometries, serve as the couplings
to the electrodes in the construction of $\boldsymbol{\Sigma}_{X}^{r}(E)$.
However, the electrode Green's functions $\boldsymbol{g}_{XX}^{r}(E)$
in Eq.~(\ref{eq:Simga_X}) are modeled as surface Green's functions
of ideal semi-infinite electrodes. In order to obtain these surface
Green's functions, we compute the electronic structure of a spherical
gold cluster with 429 atoms. From this we extract the Hamiltonian
and overlap matrix elements connecting the atom in the origin of the
cluster to all its neighbors and, using these ``bulk parameters'',
construct a semi-infinite crystal which is infinitely extended perpendicular
to the transport direction. The surface Green's functions are then
calculated from this crystal using the so-called decimation technique.\cite{Guinea:PRB1983}
We have checked that the electrode construction in the employed nonorthogonal
basis set has converged with respect to the size of the Au cluster,
from which we extract our parameters.\cite{Pauly:PhD2007} In this
way we describe the whole system consistently within DFT, using the
same nonorthogonal basis set and exchange-correlation functional everywhere.

We assume the Fermi energy $E_{F}$ to be fixed by the gold leads.
From the Au$_{429}$ cluster we obtain a Fermi energy for gold of
$E_{F}=-5.0$ eV, which is chosen to lie halfway between the levels
of the highest occupied molecular orbital (HOMO) and the lowest unoccupied
molecular orbital (LUMO) of the cluster of $-4.96$ and $-5.01$ eV,
respectively.

\subsection{Conductance}

The low-bias conductance in the Landauer formalism is given by\cite{Datta:Cambridge1995,Gosh:NanoLett2004}\begin{equation}
G_{\varphi}(T)=G_{0}\int_{-\infty}^{\infty}dE\left[-\frac{\partial}{\partial E}f(E)\right]\tau_{\varphi}(E).\label{eq:GphiT}\end{equation}
In this expression $T$ is the temperature, $G_{0}=2e^{2}/h$ is the
conductance quantum, $\tau_{\varphi}(E)$ is the energy-dependent
transmission {[}Eq.~(\ref{eq:TE}){]}, and $f(E)=1/\left[e^{-(E-E_{F})/k_{B}T}+1\right]$
is the Fermi function with Boltzmann's constant $k_{B}$. The factor
$-\partial f(E)/\partial E$ is the so-called thermal broadening function.\cite{Datta:Cambridge1995}
For zero temperature, Eq.~(\ref{eq:GphiT}) reduces to $G_{\varphi}(T=0\,\mathrm{K})=G_{0}\tau_{\varphi}(E_{F})$.
Here and henceforth we index $G$ and $\tau$ with $\varphi$, which
parameterizes different geometrical contact configurations. In our
case these configurations correspond to different molecular conformations
with the tilt angle $\varphi$ between two phenyl rings (Fig.~\ref{cap:Au-R2S2D2-Au}).
At finite temperature, tilt angles $\varphi$ differing from the minimum-energy,
ground-state value can be accessed. We account for this additional
temperature-dependent contribution to the conductance by the thermal
average\cite{Gosh:NanoLett2004,Troisi:NanoLett2004}\begin{equation}
\bar{G}(T)=\left\langle G_{\varphi}(T)\right\rangle _{\varphi}\label{eq:GT}\end{equation}
with $\left\langle \cdots\right\rangle _{\varphi}=\int d\varphi e^{-E_{\varphi}/k_{B}T}\left(\cdots\right)/\int d\varphi e^{-E_{\varphi}/k_{B}T}$.
In this expression $E_{\varphi}$ is the energy of the metal-molecule-metal
contact for angle $\varphi$. The contribution of $G_{\varphi}(T)$
to $\bar{G}(T)$ can be seen as a {}``electronic'' or {}``lead-induced''
temperature dependence, because it follows from the Fermi functions
of the electrodes. Its determination is discussed in App.~\ref{sec:App-GT}.
On the other hand, the $\varphi$ average represents a {}``configuration-induced''
contribution to $\bar{G}(T)$. For the later discussion, we also introduce
the variance \begin{equation}
\delta G(T)=\sqrt{\left\langle (G_{\varphi}(T)-\bar{G}(T))^{2}\right\rangle _{\varphi}}\label{eq:varGT}\end{equation}
that describes the fluctuations of the conductance.

\section{Results and Discussion\label{sec:Results-and-Discussion}}

Let us first discuss some properties of the isolated molecules (Fig.~\ref{cap:R2S2D2}).
For R2, S2, and D2 we obtain phenyl-ring tilt angles $\varphi$ of
$36.4°$, $90.0°$, and $90.0°$, respectively. The tilt angle of
R2 is a result of the interplay between the $\pi$ conjugation, which
tries to flatten the structure ($\varphi=0°$), and the steric repulsion
of the hydrogen atoms in the ortho positions with respect to the ring-connecting
carbons, which favors tilt angles close to $\varphi=90°$ (Ref.~\onlinecite{Pacios:ChemPhysLett2006}).
The methyl groups introduced in S2 and D2 increase the steric repulsion
and cause a larger $\varphi$. As a consequence, the conjugation between
the phenyl rings is largely broken in S2 and D2, whereas R2 still
preserves a reasonable degree of delocalization of the $\pi$-electron
system over the whole molecule. This fact is clearly reflected in
the change of the HOMO-LUMO gaps $\Delta$, which are 3.85~eV for
R2, 4.74~eV for S2, and 4.70~eV for D2. Thus, $\Delta$ increases
by roughly 1~eV when going from R2 to S2 or D2. This suggests that
the molecules S2 and D2 will show a more insulating behavior than
R2, when they are incorporated into a molecular contact.

Now, we study the geometric structure of the metal-molecule-metal
contacts. In Fig.~\ref{cap:Au-R2S2D2-Au} we show the biphenyl molecules
contacted at both ends to gold electrodes via sulfur bonds, where
the sulfur resides on the threefold hollow position of Au {[}111{]}
pyramids. The molecular conformation in the junction is very similar
to the ground-state structure of the isolated molecule, as there is
no internal stress on the molecule in this binding position.\cite{Pauly:lengthdep2007,Pauly:PhD2007}
In particular, we obtain ground-state tilt angles $\varphi_{0}$ of
$33.8°$, $89.3°$, and $89.7°$ for R2, S2, and D2, respectively. 

\begin{figure}
\begin{center}\includegraphics[%
  width=0.96\columnwidth]{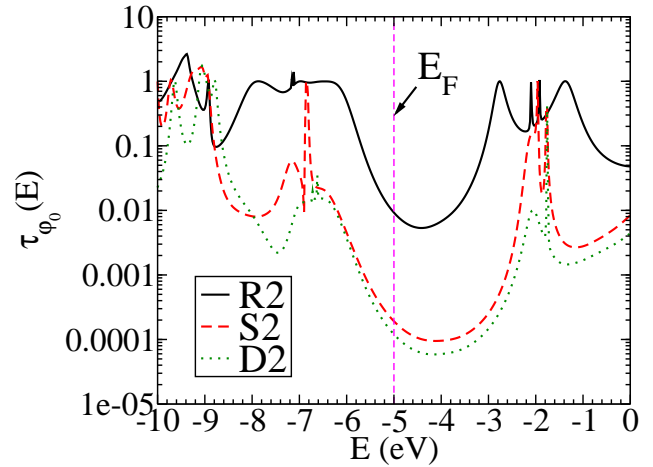}\end{center}

\caption{\label{cap:TE-R2S2D2}(Color online) Transmission $\tau_{\varphi_{0}}(E)$
as a function of energy $E$ for the ground-state geometries of the
contacts shown in Fig.~\ref{cap:Au-R2S2D2-Au}. The zero-temperature
conductances $G_{\varphi_{0}}(T=0\,\mathrm{K})$ of molecules R2,
S2, and D2 are $9.2\times10^{-3}G_{0}$, $1.9\times10^{-4}G_{0}$,
and $1.2\times10^{-4}G_{0}$, respectively. The vertical dashed line
indicates the Fermi energy $E_{F}$.}
\end{figure}

In order to analyze the conduction properties of these molecular junctions,
we have computed the transmission $\tau_{\varphi_{0}}(E)$ as a function
of energy for the ground-state geometries of the contacts (angle $\varphi_{0}$
in Fig.~\ref{cap:Au-R2S2D2-Au}). The results are plotted in Fig.~\ref{cap:TE-R2S2D2},
and our transmission curve for R2 agrees well with previous theoretical
studies.\cite{Kondo:JPhysChemA2004}%
 Obviously, $\tau_{\varphi_{0}}(E)$ is dominated by a gap around
the Fermi energy $E_{F}$, which reflects the HOMO-LUMO gaps $\Delta$
of the isolated molecules. As can be expected due to the similar geometric
conformations of molecules S2 and D2 with $\varphi_{0}\approx90°$,
their transmission curves closely resemble each other. However, the
most important observation to be made from Fig.~\ref{cap:TE-R2S2D2}
is the great reduction of the transmission $\tau_{\varphi_{0}}(E_{F})$
at the Fermi energy for S2 and D2 as compared to R2. In particular,
the conductance of S2 (D2) is lower than that of R2 by a factor of
48 (77), i.e.~roughly by two orders of magnitude. This clearly reveals
the importance of the conjugated $\pi$-electron system for the charge
transport in biphenyl molecules.\cite{Samanta:PRB1996} In addition,
it shows that the conductance can be tailored by means of adequate
side groups that force the biphenyl molecules to adopt different molecular
conformations.\cite{Venkataraman:Nature2006}

\begin{figure*}
\begin{center}\includegraphics[%
  width=2\columnwidth,
  keepaspectratio]{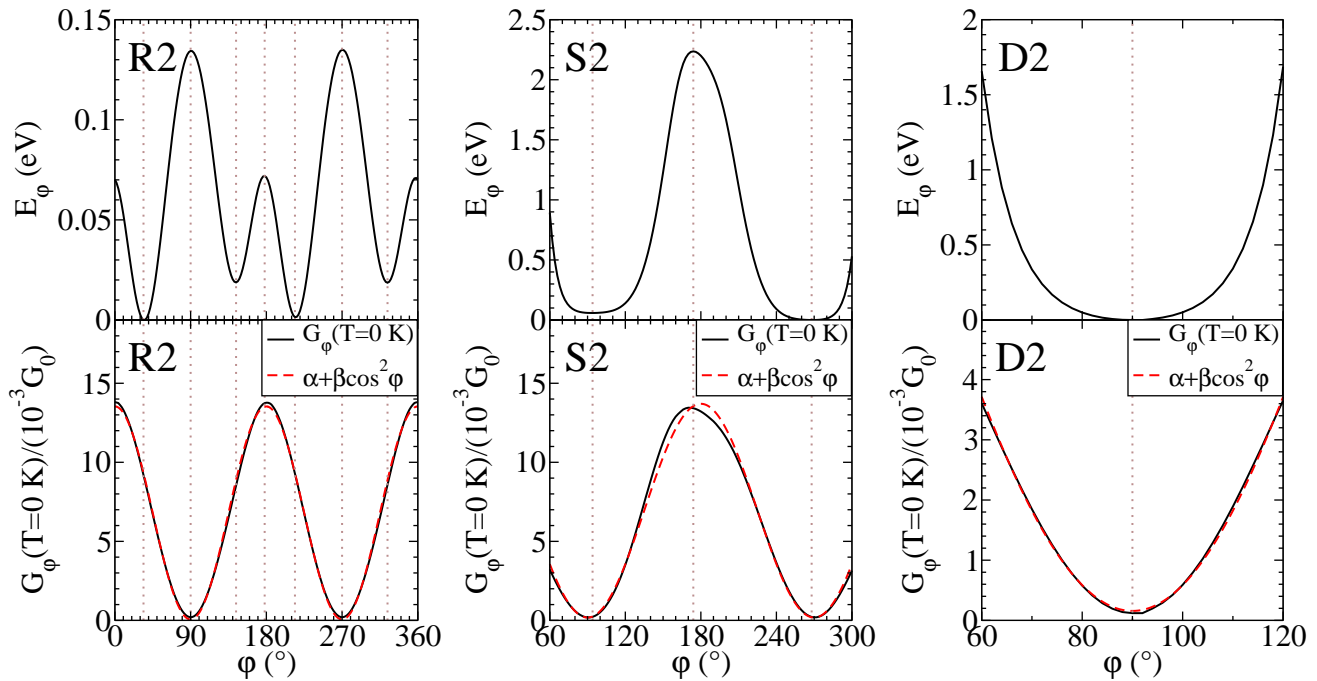}\end{center}

\caption{\label{cap:EGphi-R2S2D2}(Color online) Landscapes of the total energy
$E_{\varphi}$ (upper panels) and the conductance $G_{\varphi}(T=0\,\mathrm{K})$
(lower panels) as a function of tilt angle $\varphi$ for the molecules
R2, S2, and D2. Dotted vertical lines indicate positions of extrema
in $E_{\varphi}$ for the respective molecules (all panels). In the
lower panels a function of the form $\alpha+\beta\cos^{2}\varphi$
has been fitted to $G_{\varphi}(T=0\,\mathrm{K})$ (see the legend).
For the fit parameters we obtain $\alpha=5.95\times10^{-5}G_{0}$
and $\beta=1.35\times10^{-2}G_{0}$ (R2), $\alpha=1.81\times10^{-4}G_{0}$
and $\beta=1.35\times10^{-2}G_{0}$ (S2), and $\alpha=1.58\times10^{-4}G_{0}$
and $\beta=1.42\times10^{-2}G_{0}$ (D2). }
\end{figure*}

To investigate the dependence of the conductance on the tilt angle
in more detail, we have continuously varied $\varphi$ for all the
contacts. We do this by rotating one of the phenyl rings with respect
to the other, as illustrated in Fig.~\ref{cap:Au-R2S2D2-Au}, without
relaxing the contact geometries for tilt angles deviating from $\varphi_{0}$.
By changing $\varphi$, we obtain the results depicted in Fig.~\ref{cap:EGphi-R2S2D2},
where the total energy $E_{\varphi}$ and the conductance $G_{\varphi}(T=0\,\mathrm{K})$
are plotted as a function of $\varphi$ (Ref.~\onlinecite{noteOverestimationEphi,Almenningen:JMolStruct1985,Arulmozhiraja:JChemPhys2001}).%
 We explore the tilt-angle intervals of $\left[0°,\,360°\right[$
for R2, $\left[60°,\,300°\right]$ for S2, and $\left[60°,\,120°\right]$
for D2 (Ref.~\onlinecite{noteChoicephi}). In each case the angular
resolution is $\Delta\varphi=2°$.

In the energy curve $E_{\varphi}$ of molecule R2 there are eight
extrema visible, four minima and four maxima. They are located at
$34°$, $144°$, $214°$, and $324°$ for the energy minima and $90°$,
$178°$, $270°$, and $358°$ for the maxima. Due to the symmetry
of the biphenyl molecule, one would expect a $180°$ periodicity and
a mirror symmetry of both $E_{\varphi}$ and $G_{\varphi}(T=0\,\mathrm{K})$
with respect to $0°$ (or equivalently $90°$, $180°$, or $270°$).
While the $180°$ periodicity is present for $E_{\varphi}$, the mirror
symmetry is violated, as one can see in Fig.~\ref{cap:EGphi-R2S2D2}.
The reason for this is that the hydrogen atoms have been fixed in
their positions with respect to the phenyl rings as obtained for the
ground-state tilt angle $\varphi_{0}$. They are standing slightly
away from the phenyl ring planes in this position, which leads to
the observed violation of the mirror symmetry.\cite{Arulmozhiraja:JChemPhys2001,Sancho-Garcia:JChemPhys2004}
Contrary to $E_{\varphi}$, all expected symmetries are restored for
the conductance. In particular, $G_{\varphi}(T=0\,\mathrm{K})$ possesses
only two minima at $90°$ and $270°$ and two maxima at $0°$ and
$180°$. As a function of tilt angle, the conductance changes from
$2.0\times10^{-4}G_{0}$ in the minima to $1.4\times10^{-2}G_{0}$
in the maxima, that is, it changes by a factor of 70. 

In the case of molecules S2 and D2, $G_{\varphi}(T=0\,\mathrm{K})$
follows closely the shape of the energy curve. For S2 there are two
minima in $E_{\varphi}$ at $94°$ and $268°$ with conductances of
$2.2\times10^{-4}G_{0}$, separated by a local maximum at $174°$
with a conductance of $1.3\times10^{-2}G_{0}$. This corresponds to
a ratio of 60 between the maximum and minimum conductance. D2 exhibits
an energy minimum at $90°$ and the conductance at this point is $1.2\times10^{-4}G_{0}$. 

The close agreement of the minimal conductances for R2, S2, and D2
(Figs.~\ref{cap:TE-R2S2D2} and \ref{cap:EGphi-R2S2D2}) is remarkable.
In the conductance minima the conformations of these molecules are
the same, except for their different side groups and their slightly
varying orientations with respect to the gold electrodes. These observations
demonstrate that the side groups control the conformation, but otherwise
have little impact on the zero-temperature conductance. This is in
agreement with the experimental observations of Ref.~\onlinecite{Venkataraman:Nature2006}.

The large ratios between maximal and minimal conductances (70 for
R2 and 60 for S2) reported above highlight the relevance of the extent
of the conjugated $\pi$-electron system on the conduction properties
of the biphenyl molecules. In order to further investigate this, we
have fitted the $G_{\varphi}(T=0\,\mathrm{K})$ curves of Fig.~\ref{cap:EGphi-R2S2D2}
to functions of the form $\alpha+\beta\cos^{2}\varphi$ (see the figure
caption for the obtained fit parameters). A behavior $G_{\varphi}(T=0\,\mathrm{K})/G_{0}\propto\cos^{2}\varphi$
is expected if the coupling between the $\pi$-electron systems of
the two phenyl rings plays the dominant role in charge transport,
as discussed in more detail in App.~\ref{sec:App-eff-pi-coupl-model}.
For all three molecules our fit matches $G_{\varphi}(T=0\,\mathrm{K})$
very well. What is more, we obtain a very similar parameter $\beta$
for all of them. On the other hand, the small but positive values
of $\alpha$ indicate that the conductance at perpendicular tilt angles
($\varphi=90°$ or $270°$) does not vanish entirely, as a pure $\cos^{2}\varphi$
dependence would suggest. This observation was also made in Ref.~\onlinecite{Venkataraman:Nature2006}.
The absence of a complete blockade of the transport can be understood
by the presence of other than the $\pi$-$\pi$ couplings.

Next, we analyze the behavior of the conductance with respect to temperature.
In addition to $\bar{G}(T)=\left\langle G_{\varphi}(T)\right\rangle _{\varphi}$
{[}Eq.~(\ref{eq:GT}){]} we study $\left\langle G_{\varphi}(T=0\,\mathrm{K})\right\rangle _{\varphi}$.
In this way we can quantify the lead-induced contribution to the temperature
dependence of $\bar{G}(T)$. To perform the average $\left\langle G_{\varphi}(T=0\,\mathrm{K})\right\rangle _{\varphi}$
we use the energy and conductance landscapes $E_{\varphi}$ and $G_{\varphi}(T=0\,\mathrm{K})$
of the gold-molecule-gold contacts as shown in Fig.~\ref{cap:EGphi-R2S2D2}.
For $\bar{G}(T)$, instead, we have calculated the transmission function
for each angle in an interval around $E_{F}$ in order to obtain $G_{\varphi}(T)$
(see the explanations in Sec.~\ref{sec:Theoretical-Model} and App.~\ref{sec:App-GT}).\cite{notePhidiscretesums,noteGTintervals}

\begin{figure}
\begin{center}\includegraphics[%
  width=0.8\columnwidth,
  keepaspectratio]{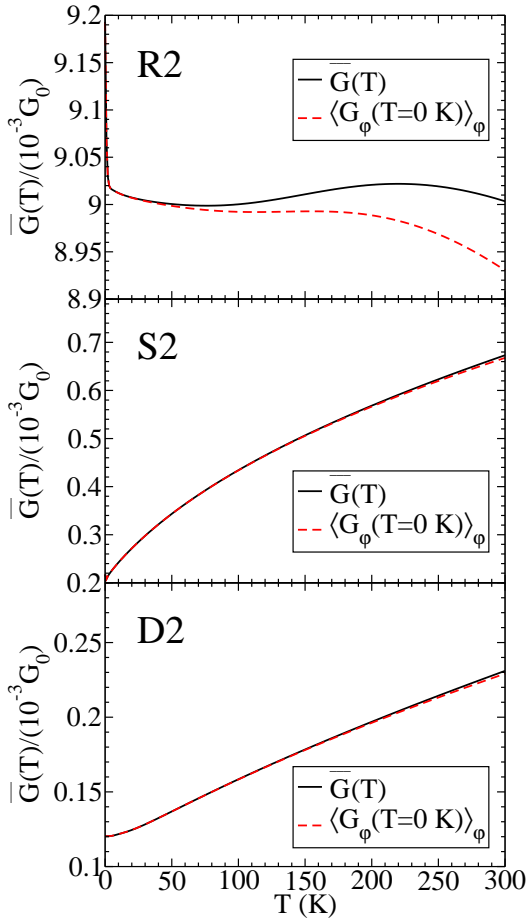}\end{center}

\caption{\label{cap:GT-R2S2D2}(Color online) Behavior of the conductances
$\bar{G}(T)$ and $\left\langle G_{\varphi}(T=0\,\mathrm{K})\right\rangle _{\varphi}$
as a function of temperature $T$ for the molecules R2, S2, and D2.}
\end{figure}

The temperature-dependent conductances $\bar{G}(T)$ and $\left\langle G_{\varphi}(T=0\,\mathrm{K})\right\rangle _{\varphi}$
are plotted in Fig.~\ref{cap:GT-R2S2D2} for the molecules R2, S2,
and D2 for temperatures $T$ between $0$ and $300$~K.%
 With respect to the behavior of $\bar{G}(T)$ we observe qualitative
differences for the three molecules considered. S2 and D2 exhibit
a monotonously increasing conductance $\bar{G}(T)$ with increasing
$T$. In contrast, after an initial drop, a non-monotonous weak temperature
dependence is found for R2.

The small differences between $\bar{G}(T)$ and $\left\langle G_{\varphi}(T=0\,\mathrm{K})\right\rangle _{\varphi}$
for S2 and D2 indicate that for these molecules the lead-induced contribution
to the temperature dependence can be neglected as compared to the
configurational one. The monotonous increase of $\bar{G}(T)$ can
therefore be understood by $E_{\varphi}$ and $G_{\varphi}(T=0\,\mathrm{K})$
(Fig.~\ref{cap:EGphi-R2S2D2}). The ground-state or equivalently
zero-temperature configurations for both molecules correspond to conformations
with minimal conductances. Therefore elevated temperatures give access
to conformations with higher conductance values, resulting in the
observed steady increase of $\bar{G}(T)$. For molecule R2 the situation
is different. Here, the energy $E_{\varphi_{0}}$ at the ground-state
tilt angle of $\varphi_{0}=34°$ does not correspond to a minimum
of $G_{\varphi}(T=0\,\mathrm{K})$. Elevated temperatures give access
to both higher and lower conductances and, as a result, $\left\langle G_{\varphi}(T=0\,\mathrm{K})\right\rangle _{\varphi}$
exhibits no clear trend. The differences between $\bar{G}(T)$ and
$\left\langle G_{\varphi}(T=0\,\mathrm{K})\right\rangle _{\varphi}$
signify that for R2 both contributions to the temperature dependence
of the conductance, namely the lead-induced and the configuration-induced
ones, play an equally important role.\cite{noteGTterms}

\begin{figure}
\begin{center}\includegraphics[%
  width=0.8\columnwidth,
  keepaspectratio]{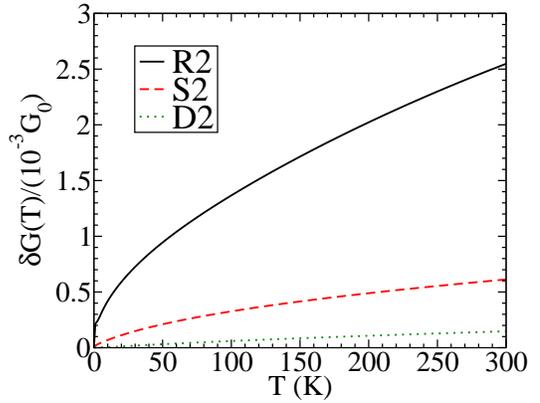}\end{center}

\caption{\label{cap:var-GT-R2S2D2}(Color online) Fluctuations $\delta G(T)$
of the conductance as a function of temperature $T$ for the molecules
R2, S2, and D2.}
\end{figure}

Finally we analyze the fluctuations of the conductance $\delta G(T)$
{[}Eq.~(\ref{eq:varGT}){]}, which we have plotted in Fig.~\ref{cap:var-GT-R2S2D2}
for temperatures $T$ ranging between 0 and 300~K.%
 In each case $\delta G(T)$ increases monotonously with $T$. This
is clear, as finite temperatures give access to conductances differing
from the ground-state conductance $G_{\varphi_{0}}(T=0\,\mathrm{K})$
(Fig.~\ref{cap:EGphi-R2S2D2}). It is also evident from Fig.~\ref{cap:var-GT-R2S2D2}
that $\delta G(T)$ is largest for R2 and smallest for D2. Indeed,
because of its two methyl side groups, D2 is the most rigid of the
three molecules with respect to ring tilts, while R2 can access a
large range of conductance values due to its shallow energy landscape
(Fig.~\ref{cap:GT-R2S2D2}). Since S2 has only one methyl side group,
it is an intermediate case. From Fig.~\ref{cap:var-GT-R2S2D2} we
obtain the temperature-fluctuation ratios $\delta G_{\mathrm{R2}}(T)/\delta G_{\mathrm{S2}}(T)=4.2$
and $\delta G_{\mathrm{S2}}(T)/\delta G_{\mathrm{D2}}(T)=4.2$ for
$T=300$~K, where $\delta G_{\mathrm{X2}}(T)$ refers to the variance
of the molecule $\mathrm{X2}$.

In experiments with molecular junctions the conductance in the last
plateau of an opening curve is generally attributed to that of a single
molecule. In practice the measured conductances are always time averages
over any fast fluctuations of the contact geometry, in particular
the internal conformations of the molecule. In terms of our definitions,
$\bar{G}(T)$ represents such a time average for a given contact realization,
while $\delta G(T)$ describes the fast fluctuations. 

The time-averaged single-molecule conductance may vary considerably
from one junction realization to another.\cite{Reichert:PRL2002,Xu:Science2003,Ulrich:JPhysChemB2006}
These variations, and hence the peak widths in conductance histograms,
can be attributed to two types of factors. The first one is due to
changes at the molecule-electrode interface and in the contact environment,
and the second one is due to modifications of the molecule's internal
geometric structure. Concerning the first contribution, the surfaces
of the metallic electrodes are atomically rough and disordered, and
the molecule binds differently to the electrodes in every realization
of the junction. \cite{Ramachandran:Science2003,Li:Nanotechnology2007,Ulrich:JPhysChemB2006}
As a result, the interface-related variability of the single-molecule
conductance is hard to control at present, although more reproducible
results can be achieved by a proper choice of the binding groups.\cite{Venkataraman:NanoLett2006}
Regarding the second point, however, the recent possibilities of chemical
synthesis allow the structure of a molecule to be designed. In order
to measure the conductance of a single molecule more reproducibly,
such a design should aim at making the molecule {}``rigid'' (for
example by means of side groups, as in the examples considered above).\cite{Venkataraman:Nature2006,Pauly:lengthdep2007,Pauly:PhD2007}
In this way the variability due to the changes in the internal structure
is reduced, because the bonded molecules stay closer to their ground-state
conformations in isolation.

In our analysis several simplifying assumptions have been made. In
particular, we have concentrated on a certain realization of a single-molecule
junction (Fig.~\ref{cap:Au-R2S2D2-Au}) and all temperature-induced
changes at the electrode-molecule interface have been neglected. Furthermore,
only one configurational degree of freedom of the molecule, the tilt
angle $\varphi$, has been taken into account, and we treated it as
a classical variable. Nevertheless, our analysis serves to illustrate
the importance of the temperature-related effects on the average conductance
and its fluctuations, and how these can be controlled by an appropriate
design of the molecules. By reducing uncertainties about the contact
geometries in this way, comparisons between theoretical and experimental
results can be made with a higher degree of confidence.

\section{Conclusions\label{sec:Conclusions}}

In conclusion, we studied the charge-transport properties of different
dithiolated biphenyl derivatives. We showed by means of density-functional-based
methods that the conduction properties of these molecules are dominated
by the degree of $\pi$-electron delocalization. A broken conjugation,
induced by side groups, was found to suppress the conductance by roughly
two orders of magnitude. By varying the tilt angle $\varphi$ between
the different phenyl rings, we observed a clear $\cos^{2}\varphi$
behavior of the zero-temperature conductance. However, the suppression
of the conductance for perpendicular ring configurations is not complete
due to the presence of other than $\pi$-$\pi$ couplings between
the phenyl rings. We showed that the methyl side groups in the biphenyl
molecules control the conformation, but they have little impact on
the zero-temperature conductance otherwise. All these findings are
in agreement with the experimental results of Ref.~\onlinecite{Venkataraman:Nature2006}. 

Based on the energy landscapes with respect to ring tilts, we also
determined the temperature dependence of the conductance. Here we
considered two different contributions. The first one originates from
the Fermi functions of the leads, while the other one is due to a
thermal average over different contact configurations. We observed
qualitatively different temperature characteristics for the well-conjugated
biphenyl molecule as compared to the molecules whose conjugation was
broken by means of methyl side groups. Furthermore, we illustrated
that an appropriate design can help to reduce temperature-induced
conductance fluctuations by stabilizing a molecule in a conformation
close to its ground-state structure in isolation. In this way uncertainties
with respect to the molecule's internal structure are reduced, and
a more reliable comparison between theoretically and experimentally
determined charge-transport properties of single-molecule junctions
seems possible.

\begin{acknowledgments}
We acknowledge stimulating discussions with Marcel Mayor and members
of the group for Theoretical Quantum Chemistry at the Universit\"at
Karlsruhe, in particular Uwe Huniar and Dmitrij Rappoport. In addition
we thank Reinhart Ahlrichs for providing us with \noun{Turbomole}.
This work was financially supported by the Helmholtz Gemeinschaft
(Contract No.~VH-NG-029), by the DFG within the CFN, and by the EU
network BIMORE (Grant No. MRTN-CT-2006-035859). We thank the INT at
the FZK for the provision of computational facilities.
\end{acknowledgments}
\appendix

\section{Temperature dependence of the conductance\label{sec:App-GT}}

In order to evaluate the temperature behavior of the conductance $G_{\varphi}(T)$
as defined in Eq.~(\ref{eq:GphiT}), we make a Sommerfeld expansion.\cite{Ashcroft:Saunders1976}
For this we compute $\tau_{\varphi}(E)$ {[}Eq.~(\ref{eq:TE}){]}
at 11 equally spaced points in the energy interval $-0.5\,\mathrm{eV}\leq E-E_{F}\leq0.5\,\mathrm{eV}$
around the Fermi energy $E_{F}$ for every tilt angle $\varphi$.
This interval is chosen large enough that the thermal broadening function\cite{Datta:Cambridge1995}\[
b(E,T)=-\frac{\partial}{\partial E}f(E)=\frac{1}{4k_{B}T}\mathrm{sech}^{2}\left(\frac{E-E_{F}}{2k_{B}T}\right)\]
has decayed to small values even for the highest temperatures considered.
(For example one gets $b(E_{F}\pm0.5\,\mathrm{eV},300\,\mathrm{K})/b(E_{F},300\,\mathrm{K})=1.59\times10^{-8}$.)
Next, we fit a polynomial of order $N=10$ to $\tau_{\varphi}(E)$
at these energy points according to \[
\tau_{\varphi}(E)=\sum_{n=0}^{N}\tau_{\varphi}^{\left(n\right)}\left(E-E_{F}\right)^{n}.\]
 With the coefficients $\tau_{\varphi}^{\left(n\right)}=\left.d^{n}\tau_{\varphi}(E)/dE^{n}\right|_{E=E_{F}}/n!$
determined from the least squares fit {[}where in particular $\tau_{\varphi}^{\left(0\right)}=\tau_{\varphi}(E_{F})${]},
the temperature dependence of $G_{\varphi}(T)$ is given as \begin{eqnarray}
G_{\varphi}(T) & = & G_{0}\left[\tau_{\varphi}^{\left(0\right)}(\varphi)+\sum_{m=1}^{\left\lfloor N/2\right\rfloor }\left(2-2^{2\left(1-m\right)}\right)\right.\label{eq:GphiT-Sommerfeld}\\
 &  & \left.\times\zeta(2m)\left(2m\right)!\tau_{\varphi}^{\left(2m\right)}\left(k_{B}T\right)^{2m}\right].\nonumber \end{eqnarray}
In this expression $\left\lfloor N/2\right\rfloor $ is the largest
integer smaller than or equal to $N/2$ and $\zeta(x)$ is the Riemann
zeta function.

\section{Effective $\pi$-orbital coupling model\label{sec:App-eff-pi-coupl-model}}

The dependence of charge transfer on the tilt angle $\varphi$ between
two phenyl rings has been inspected previously in Refs.~\onlinecite{Woitellier:ChemPhys1989,Samanta:PRB1996}.
In this appendix we discuss explicitly, how the $\cos^{2}\varphi$
behavior of the conductance can be understood based on an effective
$\pi$-orbital coupling model within the Green's function formalism. 

For this purpose we bring the transmission function $\tau_{\varphi}(E)$
{[}Eq.~(\ref{eq:TE}){]} into a slightly different form, following
Ref.~\onlinecite{Wohlthat:arXiv2007}. We assume that the $C$ part
of our contacts can be divided into two regions 1 and 2, where region
1 (2) is not coupled to the $R$ ($L$) part of the system via direct
hoppings or overlaps. Furthermore, regions 1 and 2 are connected to
each other by $\boldsymbol{t}_{12}=\boldsymbol{H}_{12}-E\boldsymbol{S}_{12}$.
Then we may write \begin{equation}
\tau_{\varphi}(E)=\mathrm{Tr}\left[\mathbf{A}_{11}\boldsymbol{T}_{12}\boldsymbol{A}_{22}\boldsymbol{T}_{21}\right],\label{eq:TE_ATAT}\end{equation}
 where $\boldsymbol{A}_{11}=i(\boldsymbol{g}_{11}^{r}-\boldsymbol{g}_{11}^{a})$
and $\boldsymbol{g}_{11}^{r}=\left[\boldsymbol{g}_{11}^{a}\right]^{\dagger}=\left[E\boldsymbol{S}_{11}-\boldsymbol{H}_{11}-\left(\boldsymbol{\Sigma}_{L}^{r}\right)_{11}\right]^{-1}$
are the spectral density and the Green's functions of region 1 in
the absence of $\boldsymbol{t}_{12}$, $\boldsymbol{T}_{12}=\boldsymbol{t}_{12}+\boldsymbol{t}_{12}\boldsymbol{G}_{21}^{r}\boldsymbol{t}_{12}$,
and $\boldsymbol{G}_{21}^{r}=\left(\boldsymbol{G}_{CC}^{r}\right)_{21}$.
Similar expressions hold for $\boldsymbol{A}_{22}$ and $\boldsymbol{T}_{21}$. 

In our case, the regions 1 (2) are made up of all atoms in the first
(second) phenyl ring plus the sulfur and three gold atoms to the left
(right) in region $C$ (Fig.~\ref{cap:Au-R2S2D2-Au}). To simplify
the discussion, we consider the electronic structure of the molecule
in the junctions as separable into $\sigma$ and $\pi$ valence electrons,
a procedure called $\pi$-electron approximation.\cite{Bishop:Dover1993}
Furthermore, we concentrate on the couplings between those $2p$ orbitals
on the ring-connecting carbon atoms, which contribute to the $\pi$-electron
system. These are oriented perpendicular to the respective phenyl
rings, and are thus rotated by the angle $\varphi$ with respect to
each other. The indices 1 and 2 then refer to these $2p$ orbitals,
and the matrices in Eq.~(\ref{eq:TE_ATAT}) become scalars. Within
an extended H\"uckel model $H_{12}$ is proportional to the overlap
$S_{12}$ (Refs.~\onlinecite{Wolfsberg:JChemPhys1952,Hoffmann:JChemPhys1963,Tinland:JMolStruct1969})
and the scalar coupling element $t_{12}(\varphi)=H_{12}(\varphi)-ES_{12}(\varphi)$
at tilt angle $\varphi$ is seen to be proportional to $\cos\varphi$. 

Because the Fermi energy of gold is located in the HOMO-LUMO gap of
the organic molecules (Fig.~\ref{cap:TE-R2S2D2}), $\boldsymbol{G}_{21}^{r}$
can be assumed to be small at $E_{F}$. Therefore $T_{12}(\varphi)\approx t_{12}(\varphi)$.
Since the $\varphi$ dependence of $A_{11}$ ($A_{22}$) can be expected
to be small, the $\cos^{2}\varphi$ behavior of the zero-temperature
conductance follows from Eqs.~(\ref{eq:GphiT}) and (\ref{eq:TE_ATAT})\[
G_{\varphi}(T=0\,\mathrm{K})=G_{0}\tau_{\varphi}(E_{F})\approx\left|t_{12}(\varphi)\right|^{2}A_{11}A_{22}.\]
 Here all energy-dependent quantities are evaluated at $E_{F}$. 

Small deviations from the $\cos^{2}\varphi$ dependence of $G_{\varphi}(T=0\,\mathrm{K})$
can be expected due to higher-order terms in the expansion of $T_{12}$
or other than the $\pi$-$\pi$ couplings in $\boldsymbol{t}_{12}$.
These include for example $\sigma$-$\sigma$ couplings of the ring-connecting
carbon atoms and next-nearest-neighbor couplings between regions 1
and 2.


\begin{thebibliography}{44}
\bibitem{noteChoicephi}The smaller tilt-angle intervals for S2 and D2 as compared with R2
are chosen to prevent the methyl groups from coming too close to each
other. For S2 we rotate the molecule such that each methyl group passes
a hydrogen atom, but we avoid the tilt angles where the two methyl
groups would meet. Accordingly, for D2 $\varphi$ is restricted even
further.
\bibitem{notePhidiscretesums}Due to our numerical grid with a finite resolution the averages in
Eqs.~(\ref{eq:GT}) and (\ref{eq:varGT}) are indeed discrete sums
over angles $\varphi_{i}=i\times\Delta\varphi$ with $\Delta\varphi=2°$.
\bibitem{noteGTintervals}We take the thermal average $\bar{G}(T)=\left\langle G_{\varphi}(T)\right\rangle _{\varphi}$
(and $\left\langle G_{\varphi}(T=0\,\mathrm{K})\right\rangle _{\varphi}$)
only over symmetry-nonredundant angle intervals, in which the ground-state
tilt angle $\varphi_{0}$ of the gold-molecule-gold contacts of Fig.~\ref{cap:Au-R2S2D2-Au}
is contained. Explicitly this means that for the molecules R2, S2,
and D2 we consider the tilt-angle intervals $\left[0°,\,90°\right]$,
$\left[180°,\,300°\right]$, and $\left[60°,\,120°\right]$ of Fig.~\ref{cap:EGphi-R2S2D2},
respectively. In this way we can get rid of the unphysical symmetry
violation of $E_{\varphi}$, visible in that figure. As explained
in the text, this symmetry violation stems the fact that we have not
performed a relaxation step for geometries deviating from the ground-state
geometry with $\varphi=\varphi_{0}$ (Fig.~\ref{cap:Au-R2S2D2-Au}).
It should be clear that the energy landscape $E_{\varphi}$ changes
upon inclusion of such a relaxation step.\cite{noteOverestimationEphi}
\bibitem{noteGTterms}For R2 the terms up to $m=2$ need to be included in the Sommerfeld
expansion of $G_{\varphi}(T)$ {[}Eq.~(\ref{eq:GphiT-Sommerfeld}){]},
in order to describe the behavior of $\bar{G}(T)=\left\langle G_{\varphi}(T)\right\rangle _{\varphi}$
in Fig.~\ref{cap:GT-R2S2D2}. In contrast, for S2 and D2 already
the $m=1$ term is almost negligible as indicated by the close agreement
of $\left\langle G_{\varphi}(T=0\,\mathrm{K})\right\rangle _{\varphi}$
and $\bar{G}(T)$.
\end{thebibliography}
\end{document}